\documentstyle[11pt,newpasp,twoside,epsf]{article}
\newcommand{\etal}{et~al.}

\newcommand{\kms}{$\mbox{km~s}^{-1}$}

\def\edcomment#1{\iffalse\marginpar{\raggedright\sl#1\/}\else\relax\fi}
\reversemarginpar

\begin{document}
\title{Detection of new sources of 4765-MHz OH masing}
\author{R. Dodson \& S Ellingsen}
\affil{School of Mathematics and Physics, University of Tasmania, 
     GPO Box 252-21, Hobart, Tasmania 7001, Australia}

\begin{abstract}
  We have used the Australia Telescope Compact Array (ATCA) to make a
  sensitive search for maser emission from the 4765-MHz transition of
  OH towards a sample of 55 star formation regions.  Maser emission
  with peak flux densities in excess of 100~mJy were detected in 14
  sites, with 10 of these being new discoveries. Unlike the
  ground-state OH transitions the 4765-MHz transition is not predicted
  to be circularly polarised and none of the masers observed have
  detectable levels of linear, or circular polarisation. Combining our
  results with those of previous high resolution observations of other
  OH transitions we are able to investigate various theoretical
  models for the pumping of OH masers.
\end{abstract}

\section{Introduction}

Molecular maser emission from the OH molecule has been found to occur
in a variety of astrophysical environments, including star formation
regions, late-type stars, SNR and the nuclei of distant
galaxies. Despite its apparent simplicity the rotational spectrum of
the hydroxl radical is complicated by $\Lambda$-doubling and hyperfine
splitting of the levels.  Towards star formation regions OH maser
emission is most frequently observed in the main-lines of the ground
state $^2\Pi_{3/2} (J=3/2)$ transitions, in particular the 1665-MHz
($F=1\rightarrow 1$) transition.  Masing in the other ground state
main-line transition at 1667~MHz ($F=2\rightarrow 2$) frequently
accompanies the 1665-MHz masers, and in some sources the 1612- or
1720-MHz satellite-line transitions also exhibit maser action.
Emission has been detected towards star formation regions for all
groups of OH transitions with energies less than approximately 500~K
above the ground state.


The higher excited OH transitions of 4750 and 6035~MHz ($^2\Pi_{1/2}$,
F=1$\rightarrow$1, $^2\Pi_{3/2}$, F=3$\rightarrow$3) (plus their
satellites) have also been observed. We have investigated the
association of the strongest $^2\Pi_{1/2}$ transition, (F=1$\rightarrow$0 at
4765.562~MHz) with other lines also observed using the
ATCA (Caswell 1997, 1998, 1999).

Several previous studies of maser sites have found that the regions of
1720-MHz also have 4765-MHz (Palmer \etal\ 1984, Masher \etal\ 1994,
MacLeod 1997).
Theoretical models of Gray \etal\ (1992) of maser pumping were
developed and found that in regions of higher temperature and medium
velocity shift have both strong emission at 1720 and 4765~MHz and
6035~MHz is supressed. The more recent modeling of Pavlakis \etal\
(1996) do not find any overlap of the 1720 and 4765 masers, yet a
common overlap of 6035 and 4765~MHz. Their calculations of the 5cm
lines are hampered by the fact that some of the theoretical collision
rate coefficients have not been calculated. Our work tests the
predictions of these models.


\section{Observations and Data reduction} 

All observations were made with the Australia Telescope Compact Array.
The configuration of six antennas yielded 15 baselines between 5 and
95 k$\lambda$. All pointings had 30 minutes or more of on-source
observation, and the primary (and bandpass) calibration was done
against PKS 1934-638.  The correlator provided a 1024 channel spectrum
with all four polarisation products across a 4-MHz bandwidth.  The
sources were selected from those observed by the ATCA to have 6035-MHz
emission (Caswell 1997), or 1720~MHz (Caswell 1999), or those
previously reported to have 4765-MHz emission. The observations of the
4765-MHz maser were made in September 2000.

\section{Results}

Data reduction was done with {\bf miriad} and {\bf karma} following the
standard methods. With 1024 channels across 4~MHz the effecitive
velocity resolution is 0.3 \kms. The 1~$\sigma$ level over this
bandwidth is $\approx$ 30 ~mJy per channel, or 14~mJy per \kms. The
positional accuracy is typically (in comparison with the reference
postions) 0.6 arcsecs. We searched an ares of 256 arcseconds centred on
the quoted site positions, a velocity range of $\pm$5 \kms\ at 0.2
\kms\ resolution.  Furthermore we searched over 150 arcseconds and
$\pm$20 \kms\ at a resolution of 1~\kms. The entire bandpass was
searched at the given position.

\begin{table}
\caption{Sources with detected 4765-MHz OH emission.  References: 
           *=new source, 
           a=Cohen, Masheder \& Caswell 1995;
           b=Gardner \& Ribes 1971;
           c=Smits 1997;
           d=Zuckerman \& Palmer 1970.}
  \label{tab:det4765}
  \begin{tabular}{lcccrcrl}
    \hline
               &  {\bf Peak Flux} & 
    {\bf Velocity} & {\bf Width } &                   \\
  {\bf Source} & {\bf Density}   &
    {\bf of Peak}  & {\bf of Peak}    &                   \\
  {\bf Name}   & {\bf (Jy)}      &
    {\bf (\kms)}   & {\bf (\kms)}   &  {\bf References} \\ \hline
  G240.316+0.071      &0.31,0.18&65.2,63.0&  0.4,0.4 &   * \\ 
  G240.31+0.07        &0.11&66.7&      0.4 &   * \\ 
  G294.511-1.621      &  2.12 & -12.0 & 0.4      &   c \\
  G309.921+0.479      &  0.17 & -61.0 & 1.8 &   * \\ 
  G328.307+0.430      & 0.21 & -90.6 &  0.5    &   * \\ %
  G328.304+0.436      & 0.13 & -92.0 &  0.3   &   * \\ 
  G328.808+0.633      &  0.16 & -44.8 & 0.5 &   * \\ 
  G328.809+0.633      &  0.13 & -43.5 & 0.5 &   * \\ 
  G333.135-0.431      &  0.14 & -51.4 & 0.4 &   * \\ 
  G333.135-0.431s     &  0.07 & -54.3 & 6.7 &   *,Thermal \\
  G353.410-0.360      & 1.68 & -20.9 &  0.4      &   a \\
  Sgr B2/G0.666-0.035 & 0.05 &  49.4 &    5.5 &   b, Thermal \\
  G011.904-0.141    & 1.0 &  41.6 &    0.4 &   * \\ 
  W49SW              &  0.24,0.12 &   8.4,11.7 &  0.4,0.2 &  a\\
  W49N               &  0.65 &  2.1 &     0.5      &   d \\
  W49NW             &  0.32 &   2.5 &     0.5      &   * \\ %
\hline
  \end{tabular}
\end{table}

Fourteen sites of 4765-MHz maser emission were detected, two
further sites showed signs of broader thermal emission, or possibly
blended maser emission. These are listed in table 1.
Ten of the sources are new detections, the remaining previously reported
in various papers, as indicated in the table.

Those sites for which we have non-detections include some which have
been reported as showing emission previously. The most likely reason for the
absence is that the 4765-MHz maser is known to be highly
variable as reported, for example, by Smits (1997).

The ratio of detections to sites with 6035-MHz emission was 8 out of
38, while for the 1720-MHz it was 2 out of 12, both of which also have
6035-MHz emission. This draws into doubt the claimed predictive power
of 1720-MHz, though with few detections the statistics are not
compelling.

\begin{figure}
        \plotone{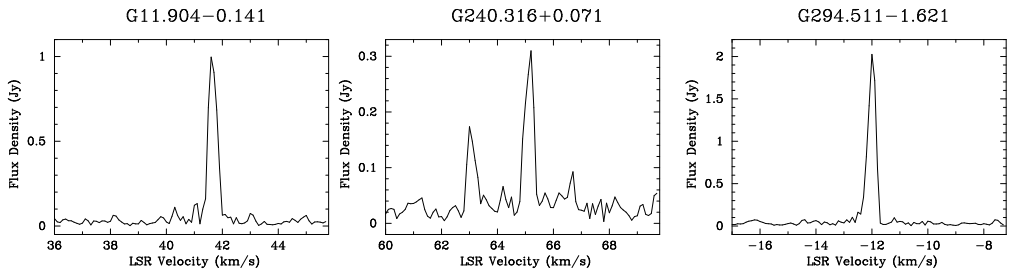}
\caption{Spectra of some of the 4765~MHz OH masers observed with the ATCA}
\label{fig:spec4765}
\end{figure}

\section{Discussion} 

The theoretical mechanisms are complicated by some unknown or poorly
known excitation rates for the pumping paths. The models of of Gray et
al.  have a well populated combination of 1720 and 4765~MHz with local
temperatures of 125K, high densities and large velocity gradients.
These are the only conditions under which 4765~MHz is significantly
excited. The 6035-MHz line is strongly supressed under these
conditions, which is in contrary to our observations in which 4765- and
6035-MHz masers are commonly present together. The more recent work of
Pavlakis \etal \ find 4765~MHz only excited with {\em low} velocity
gradients, while 1720~MHz is only exited with {\em high} velocity
gradients, in contridiction with that seen in W3(OH) (although most
observational papers draw attention to the fact this could easily be a
projection effect). They have many more free parameters in their
modelling, which complicates the intereptation, as it possible that
there is some, unpublished combination that does excite 4765, with 1720
and/or 6035~MHz.

Ideally we would like greater resolution than the 2 arc seconds
achieved by the ATCA, as this is not sufficient to separate what could
be independant masing regions powered by the same central source.

\section{Conculsions} 

We report the discovery of ten new $^2\Pi_{1/2}$ maser emission
sites, more than doubling the number reported in the southern hemisphere.

We have found that the corespondance of 1720~MHz to 4765~MHz is not as
strong as the corespondance of 6035~MHz to 4765~MHz.


\end{document}